\documentclass{llncs}
\usepackage{multirow}
\usepackage{amsmath}
\usepackage{graphicx}
\usepackage{float}
\usepackage{hyperref}
\usepackage[table]{xcolor}
\usepackage[most]{tcolorbox}
\usepackage[
backend=biber,
style=ieee,
]{biblatex}
\title{\vspace*{\fill}Automated GI tract segmentation using deep learning}
\subtitle{Track organs in medical scans to improve cancer treatment}

\author{Manhar Sharma \small{0000-0002-7839-4711} \\
Bristol, United Kingdom\\
\email{manharsharma007@gmail.com}
\vspace*{\fill}}

\begin{document}

\maketitle

\newpage

\begin{abstract}
The job of Radiation oncologists is to deliver x-ray beams pointed towards the tumor and at the same time avoid the stomach and intestines. With newer technologies such as MR-Linacs, oncologists can visualize the position of the tumor and allow for precise dose according to tumor cell presence which can vary from day to day. The current job of outlining the position of the stomach and intestines to adjust the X-ray beam’s direction for the dose delivery to the tumor while avoiding the organs. This is a time-consuming and labor-intensive process that can easily prolong treatments from 15 minutes to an hour a day unless deep learning methods can automate the segmentation process.
\\
This paper studies semantic segmentation on the GI Tract scans using deep learning to make this process faster and allow more patients to get effective treatment.

\end{abstract}

\section{Introduction}

In 2019, an estimated 5 million people were diagnosed with cancer of the gastro-intestinal tract worldwide \cite{rawla2019epidemiology}.
radiation therapy (RT) has the potential to improve the rates of cure of 3.5 million people and provide palliative relief for an additional 3.5 million people \cite{jaffray2015radiation}.
\\
The Radiation oncologists deliver x-ray beams pointed toward the tumor and at the same time avoid the stomach and intestines. With MR-Linacs (magnetic resonance imaging and linear accelerator systems) \cite{lagendijk2014magnetic}, oncologists can visualize the position of the tumor and monitor for precise dose according to tumor cell presence which can vary from day to day. The current job is to manually outline the position of the stomach and intestines for adjustments to the X-ray beam’s direction to increase the dose delivery to the tumor while avoiding the organs. This is a time-consuming and labor-intensive process that can easily prolong treatments from 15 minutes to an hour a day unless deep learning methods can be applied and could help automate the segmentation process.

Deep learning can help by automating the segmentation process to reduce manual labor and help more patients to get effective treatment. In this paper, we study the segmentation task on GI Tract scans using different backbones for the UNet architecture. Our best model yielded an IoU score of 85.3\% using the Efficientnet backbone.

\section{Related work}
Many researchers have used different deep learning architectures in medical imaging to create both semantic and instance segmentation and achieved excellent results. For instance, in medical images, deep learning approaches have been applied for brain tumor segmentation \cite{havaei2017brain}, colon polyp segmentation \cite{duc2022colonformer} and pancreas segmentation \cite{oktay2018attention}. In 2016, V-Net \cite{milletari2016v} network architecture was proposed with the aim of segmenting volumetric data such as 3D MRI scans. This approach was similar to U-Net network architecture which was proposed in 2015 \cite{ronneberger2015u} for biomedical image segmentation. U-Net relied on the strong use of data augmentation due to the fact that data in biomedical applications is not present in abundance. The effectiveness of U-Net architecture has led to the development of different architectures like U-Net++ \cite{zhou2018unet} and TMD-Net \cite{tran2021tmd} which have reported significant performance gains over U-Net in different biomedical segmentation tasks.
\\
The aim of this study is to use the existing knowledge and apply it to the problem of GI tract segmentation. There is a lack of research done on the segmentation of GI tract organs. This study aims to provide a new baseline using different backbones of U-Net architectures in order to segment organs such as stomach, large Bowel and small Bowel (Fig. 1). The task of this research is to find the best architecture to create masks for a scan that highlights the Large Bowel, Small Bowel, and Stomach. Generally, the process of applying deep learning to such tasks starts with collecting the data and effectively pre-prossessing the data to be fed into the model to create useful predictions. The next section discusses the dataset used in this research.
\begin{figure}[H]
\centering
     \includegraphics[width=1.0\textwidth]{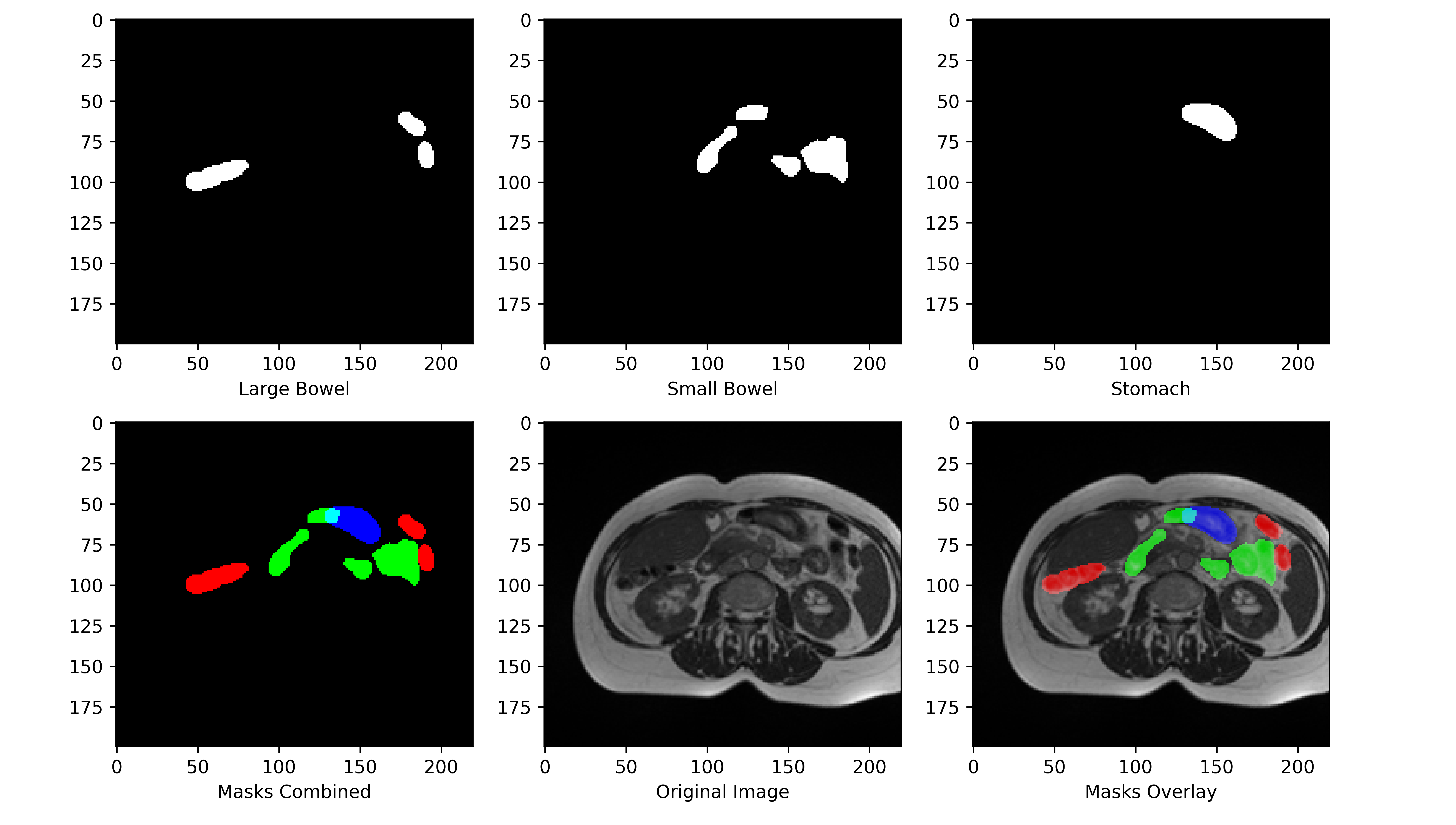}
      \caption{Dataset (scans with masks of organs)}
       \label{normal_case}
\end{figure}

\section{Dataset and Preprocessing}

The dataset used in this research is available on the Kaggle platform for free. This dataset uses anonymized scans of patients provided by the UWMadison Carbone Cancer Center. The recent competition "UW-Madison GI Tract Image Segmentation” hosted on Kaggle challenges researchers to apply Deep learning models to perform semantic segmentation of the GI tract scans. These scans are taken over a period of six days and the masks for training are provided as Run-length encoding to save memory and bandwidth. The input images are the  float16 PNG slices of the scans. 
\\
The problem with these scans is the inconsistent widths and heights (Fig. 3). Few images are square and while others are rectangular. To match the input shape of the model (288x288), zero padding is done where the length is less than 288, and pixel trimming was done where the length is greater than 288. The cropping was done in such a way that the black/empty pixels were trimmed first. Same pre-processing was done for the masks as well. 
\begin{figure}[ht]
\centering
     \includegraphics[width=6.5cm]{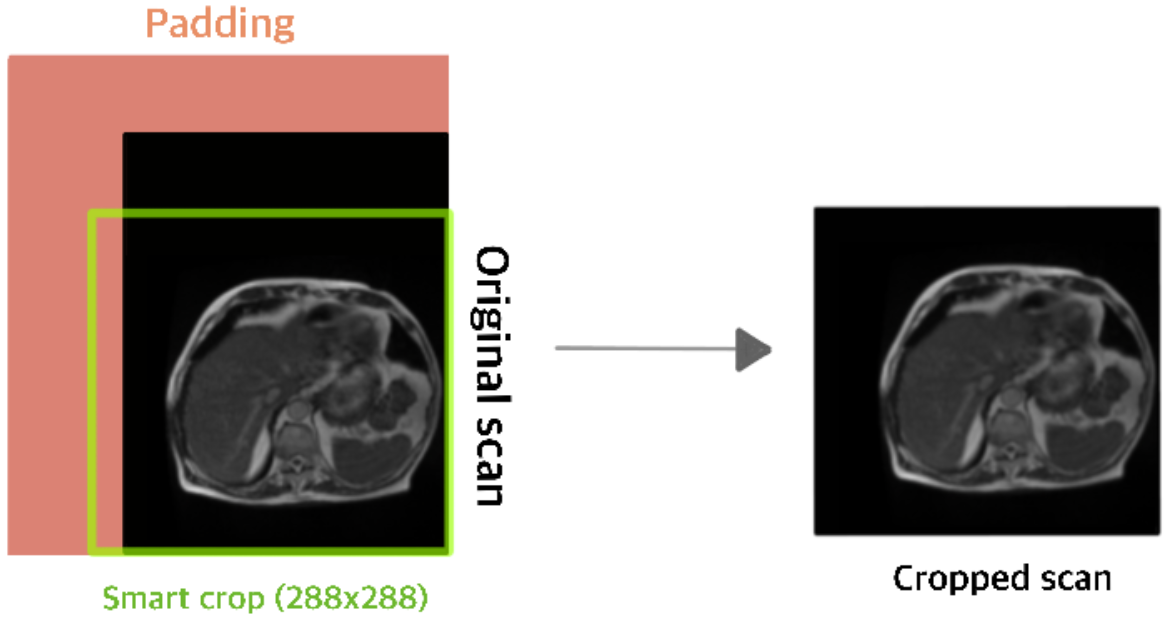}
      \caption{Rectangle/smaller scans to (288x288)}
       \label{normal_case}
\end{figure}

Resizing was not considered as this can create abnormal artifacts which can hinder the correct mask of the organ (this is crucial as the organs can be very close to each other).
\\
The same padding step will be done for the test image lower than 288x288 and then the sliding window patches of 288x288 will be generated for each image for images larger in size to make patch-wise predictions. The predicted patches will be stitched to get the image mask.
The input images were converted into tensors before being fed to the network and each image was also normalized to help the model in learning the patterns \cite{huang2020normalization}.

\begin{figure}[ht]
\centering
     \includegraphics[width=1.0\textwidth]{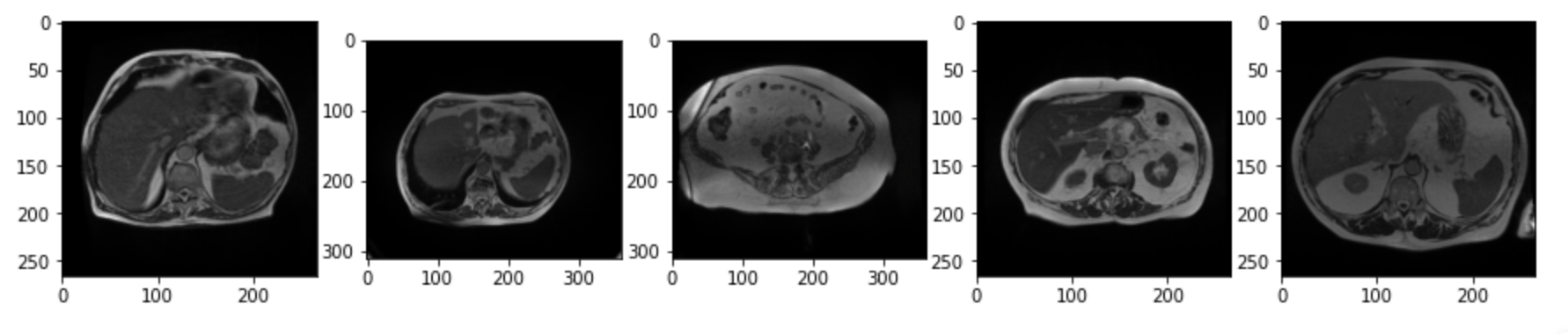}
      \caption{Shows imbalanced scans dimensions}
       \label{normal_case}
\end{figure}

\section{Structure of Model}
A convolutional neural network or CNN consists of a stack of three main neural layers: convolutional layer, pooling layer, and fully connected layer \cite{malhotra2019computer}\cite{dargan2020survey}. \\
The U-NET is a deep learning convolutional network that create segmentation masks from an input image. The U-Net consist of an encoder part that converts $n_x$-channel image to a dense encoding and a decoder part that converts the dense encoding back to the $n_y$-channel output \cite{ronneberger2015u}.
\\
In this paper, we used different encoders (ResNet\cite{resnet}, EfficientNet \cite{tan2019efficientnet}, VGG16 \cite{vgg16}, MobileNet \cite{mobilenet}) for U-Net having 1 input channel and 3 output channel and compare the results. Each of the 3 output channels corresponds to a different segmentation mask. Each channel in the output tensor is a mask for one of the three classes (Large Bowl, Small Bowel, and Stomach).
\\~\\
The encoder part of the U-Net classifies each pixel in the image into a particular class. The information is compressed using a down-sampling (max-pooling) approach and the resulting segmentation masks are reconstructed by using up-sampling (Up convolution/Convolution transpose) instead of pooling layers to improve the resolution of the output. A successive convolution layer in the decoder part can then learn to assemble a more precise output based on this information \cite{ronneberger2015u}.

\begin{figure}[ht]
\centering
     \includegraphics[width=1.0\textwidth]{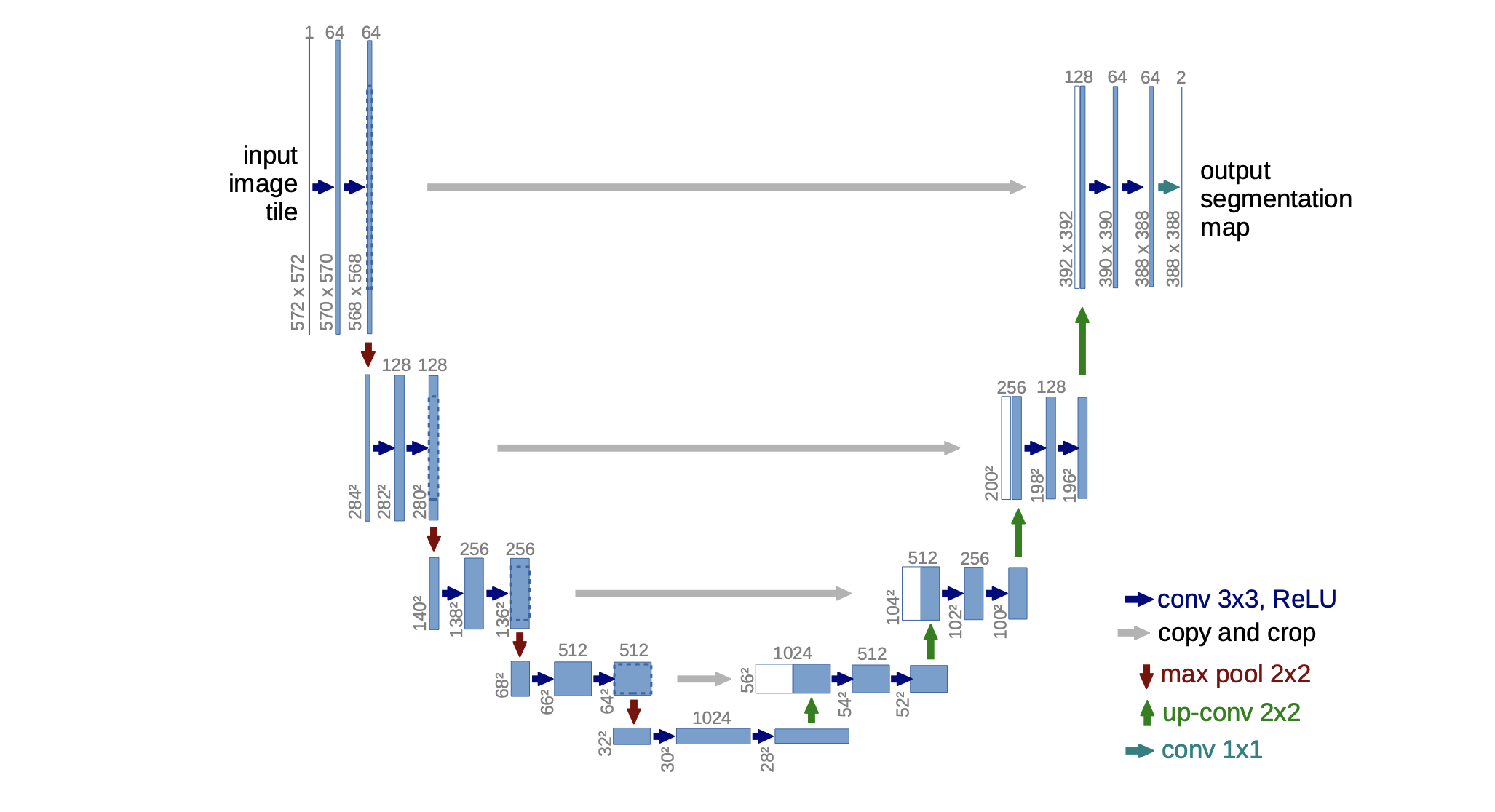}
      \caption{U-Net architecture taken from original paper \cite{ronneberger2015u}}
       \label{normal_case}
\end{figure}

\section{Performance Metrics}
There are many performance metrics that can be applied to segmentation models to measure their prediction performance. The performance metric used in this research for measuring accuracy is "Intersection over Union" and the loss functions are derived from the combinations of "Intersection over Union Loss \cite{IoULoss}", "Binary Cross Entropy Loss", "Tversky Loss \cite{tversky}".
\\~\\
\textbf{Performance Metric (For single sample) : }
    \begin{gather}
    \text{IoU}(A,B)=\frac{A \cap B}{A \cup B}
    \end{gather}
\\
\textbf{Loss functions used :}
    \begin{gather}
    \text{BCE Loss} + \text{Tversky Loss} = 0.4 * (\text{Tversky Loss}) + 0.6 * (\text{BCE Loss})
    \\
    \text{IoULoss} + \text{Tversky Loss} = 0.4 * (\text{Tversky Loss}) + 0.6 * (1 - \text{IoU})
    \\
    \text{IoULoss} = 1- \text{IoU}
    \end{gather}

\section{Training}
The input images along with their segmentation masks are used to train the network with the Adam optimizer \cite{adam} implementation of PyTorch with initial learning rate of $5e^{-3}$. Consine Annealing \cite{cascheduler} was used to decay the learning rate as after 30 epochs the model loss was not decreasing and was fluctuating within the $e$ range. To minimize the overhead and make maximum use of the GPU memory, we used 288x288 image for input with borders cropped as the borders held no information for training. For similar reasons the batch size of 32 was used and the models were trained for 80 epochs. The dataset was divided into training data and validation data using 80/20 split. The models were trained using training data and the performance was evaluated using validation data.

The output of the model was passed through a sigmoid unit and the output was clamped to 0 for values less than 0.5 and 1 otherwise. 

$$
\sigma(x)=\begin{cases}
			1, & \text{if $x$ $\geq$ 0.5}\\
            0, & \text{otherwise}
		 \end{cases}
$$

\section{Experimentation Results}
The results of the experimentations are very clear from the table 1. The Efficientnet encoders  outperformed all the other other encoders. The models were trained for 80 epochs. No data augmentation other than Horizontal flipping and vertical flipping were used. Different augmentation techniques like ElasticTransform and ShiftScaleRotate were tried but they had no visible effect on the performance of the model.
Although VGG16 was the least performing encoder among other encoders, yet it managed to achieve over 80\% accuracy using (Tversky + BCE) Loss function.
Upon experimentations it is clear that the best choice of loss function for this task is (BCE + Tversky) Loss (Fig. 7).

\begin{figure}[H]
\centering
     \includegraphics[width=1.0\textwidth]{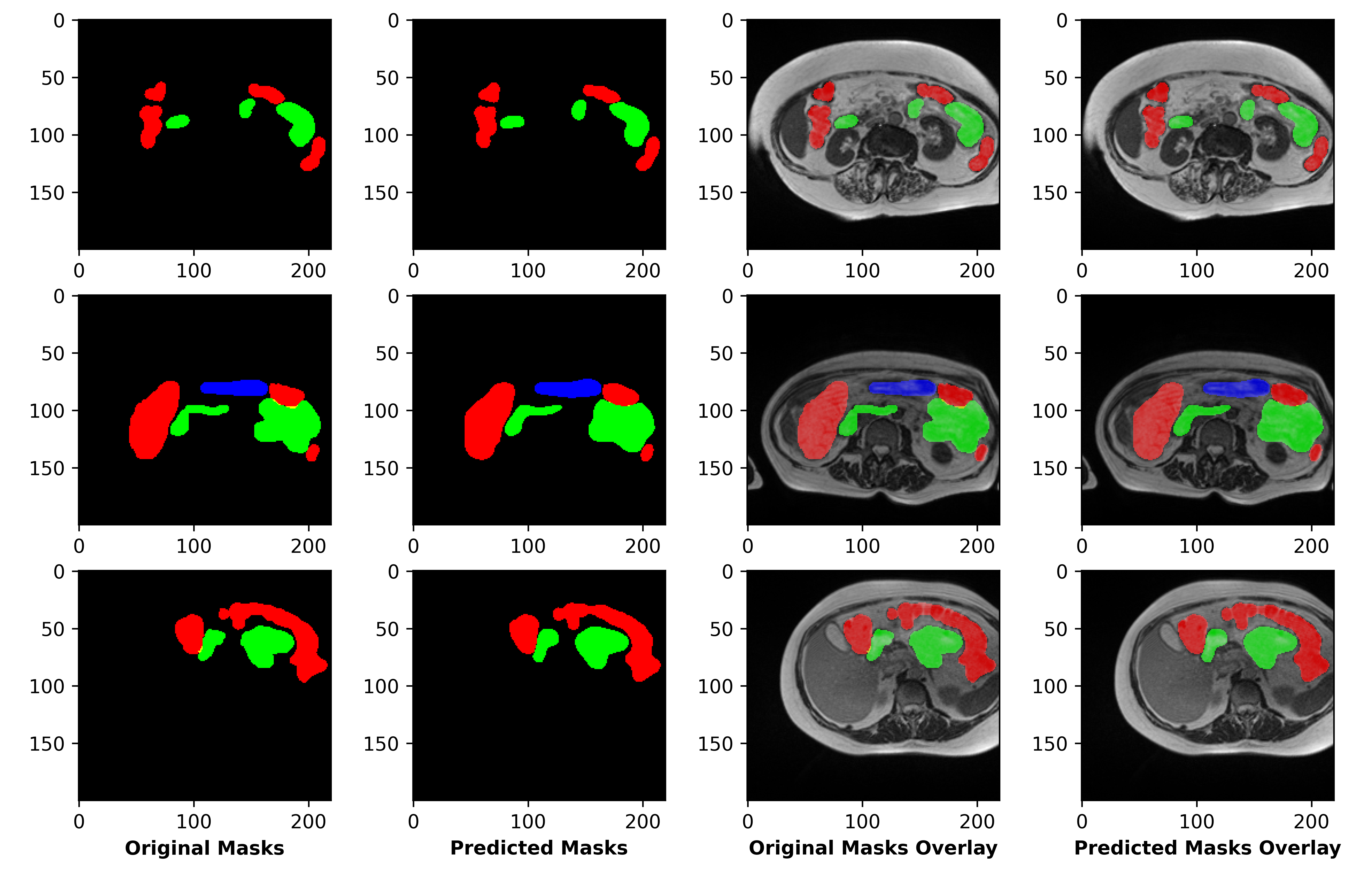}
      \caption{Final Predictions after training}
       \label{normal_case}
\end{figure}

\setlength{\tabcolsep}{7.5pt}
\renewcommand{\arraystretch}{1.2}
\begin{table}[hbt!]
\centering
    \begin{tabular}{ |p{2.3cm}|p{1.8cm}|p{3cm}|p{3cm}|  }
            \hline
            \rowcolor{lightgray} \multicolumn{4}{|c|}{Validation Accuracy over after 80 Epochs} \\
            \hline
            Encoder & IoU Loss & BCE + Tversky Loss & IoU + Tversky Loss \\
            \hline
            Efficientnet-B3 & 84.9\% & 85.3\% & 84.8\% \\
            Efficientnet-B1 & 84.3\% & 85\% & 84.5\% \\
            Resnet34 & 84.6\% & 85.2\% & 83.9\% \\
            Resnet50 & 83.7\% & 84.9\% & 83.5\% \\
            Mobilenet V2 & 81.6\% & 84.1\% & 83.7\% \\
            VGG16 & 79.5\% & 82.7\% & 65.7\% \\
            \hline
    \end{tabular}
\vspace{0.2 cm}
\caption{\label{tab:table-name}List of encoders used along with the results achieved}
\end{table}

\begin{figure}[H]
\centering
     \includegraphics[width=1.0\textwidth]{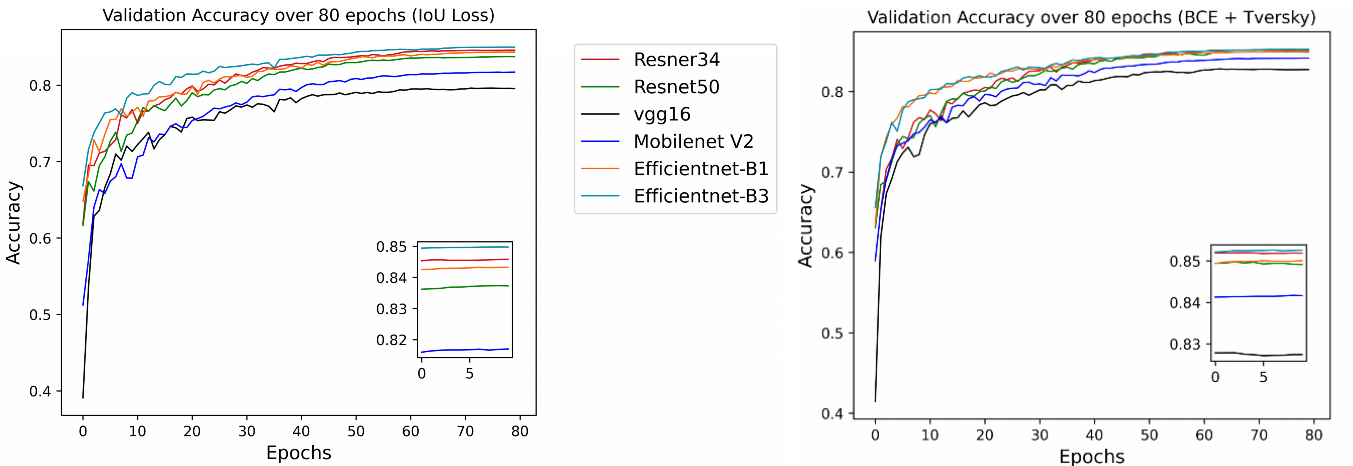}
      \caption{Accuracy achieved using IoU and (Tversky + BCE) Loss functions}
       \label{normal_case}
\end{figure}

\begin{figure}[H]
\centering
     \includegraphics[width=1.0\textwidth]{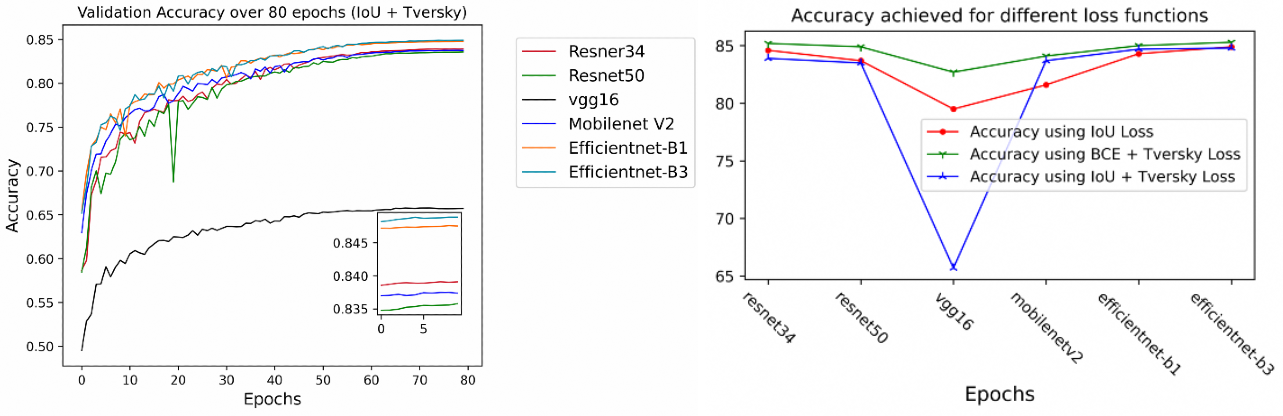}
      \caption{Accuracy achieved using (Tversky + IoU) Loss function and BCE+Tversky being the best Loss Function among IoU, IoU+Tversky, BCE+Tversky}
       \label{normal_case}
\end{figure}

\section{Conclusion}
This research used classic U-Net architecture with different encoder. There are more advanced algorithms available that have achieved excellent results in different classification tasks. These algorithms can be applied as an encoder to create a new U-Net flavor and achieve better results. Though the loss functions used in this research are IoU Loss, (BCE + Tversky) Loss, (IoU + Tversky) Loss, different loss functions can be applied to experiment with the results. A few of them are Binary cross entropy Loss, Dice Loss, and Focal Loss (For imbalanced masks) and combination of these to create new loss functions. This research can serve as a baseline for new research in the same criteria. Also, scans for the same patients on the same day can be stacked to make a high channel input and predict the mask based on the $n+1^{th}$ channel. $0-n$ channels can serve as past history for the $n+1^{th}$ prediction.

\section{Acknowledgements}
This study was made possible with the anonymized data provided by The UW-Madison Carbone Cancer Center on Kaggle platform.
This research was made possible by the repository "Segmentation models.PyTorch" which provides a high level API for different encoder implementation for UNet \cite{modelsPy}.

\section{Data availability}
Data used in this research is available on kaggle platform under the competition ”UW-Madison GI Tract Image Segmentation” and can be downloaded from \href{https://www.kaggle.com/competitions/uw-madison-gi-tract-image-segmentation/data}{https://www.kaggle.com/competitions/uw-madison-gi-tract-image-segmentation}.

\section{Statements \& Declarations}
\subsection{Funding}
The author declare that no funds, grants, or other support were received during the preparation of this manuscript.

\subsection{Competing Interests}
The author has no relevant financial or non-financial interests to disclose.

\subsection{Author Contributions}
Material preparation, data collection and analysis was performed by Manhar Sharma. The author read and approved the final manuscript.

\printbibliography

\end{document}